\documentstyle[11pt,newpasp,twoside,epsf]{article}
\def\edcomment#1{\iffalse\marginpar{\raggedright\sl#1\/}\else\relax\fi}
\reversemarginpar

\begin{document}
\title{DENIS Survey Data Probing Different Types of PNe}
 \author{Stefan Schmeja \& Stefan Kimeswenger}
\affil{Institut f\"ur Astrophysik, Leopold-Franzens-Universit\"at Innsbruck, Technikerstr. 25, A-6020 Innsbruck, Austria}

\begin{abstract}
We present near-infrared photometry of a large sample of PNe and show that the {\it IJK} colours are a good tool to distinguish different classes of PNe, like nebulae around symbiotic Miras or IR-[WC] PNe from genuine PNe.
\end{abstract}

\noindent Near infrared (NIR) photometry of planetary nebulae (PNe) allows a classification of those objects (Whitelock 1985). We present a sample of 137 PNe observed with DENIS (Deep Near Infrared Southern Sky Survey; Epchtein et al.\ 1997) in Gunn-$I$ (0.82~\micron), $J$ (1.25~\micron), and $K{\rm _s}$ (2.15~\micron). 
Details on the sample and on the data reduction are given in Schmeja \& Kimeswenger (2001a).

Genuine PNe are located in a well defined region in the dereddened $IJK$ two-colour diagram (Fig.~1).
Several classes of PNe, however, show colours different from genuine PNe. While all the objects lie in the region of genuine PNe in optical classification diagrams, the NIR colours differ. 

{\it Nebulae around symbiotic Miras} (Corradi et al.\ 1999) look very much like PNe, although they are formed in a slightly different way. 
In the two-colour diagram the known and suspected symbiotic Miras are pretty well separated from the genuine PNe as well as from the classical Miras. Thus our diagram is a unique tool to distinguish genuine from symbiotic PNe. 
For a detailed analysis see Schmeja \& Kimeswenger (2001b).

{\it Infrared [WC] stars} are a separate subclass of objects belonging to the Wolf-Rayet stars (Zijlstra 2001). While normal [WC] PNe have the same IRAS colours as other PNe, the IR-[WC] PNe have unusual colours resembling young post-AGB stars. In our diagram they show usual $(I-J)_0$ colours but higher $(J-K)_0$ values compared to the other PNe.
Polycyclic aromatic hydrocarbon (PAH) features were detected between 3 and 12~\micron\ in ISO SWS spectra of He~3-1333 (PN~G332.9-09.9), He~2-113 
(PN~G321.0+03.9), and Vo~1 (PN~G291.3-26.2) (Szczerba et al.\ 2001). They may increase the flux in $K$ leading to the unusual high $(J-K)_0$ value.

{\it Peculiar Objects:}
PM~1-188 (PN~G012.2+04.9) has a very late [WC] type central star and a very low density for its spectral type. Pe\~{n}a, Stasi\'{n}ska, \& Medina (2001) therefore conclude that it either underwent a late helium flash (``born again'' scenario) or that it evolved particularly slowly off the AGB. In our diagram PM~1-188 has a usual $(I-J)_0$ but a very high $(J-K)_0$ value, similar to the known ``born again'' PN A58 (V605~Aql). Thus the ``born again'' scenario may in fact be the case for PM~1-188, too.\\
SB~17 (V348~Sgr, PN~G011.1-07.9) is a nebula whose true nature is still not clear: 
Apart from a PN it is regarded as a [WC]~PN (Koesterke 2001) or an R Coronae Borealis star (Clayton 2001). 
Clayton (2001) claims that a relation exists between the two latter types and mentions V348~Sgr and He~3-1333 as examples showing characteristics of both types. In our $IJK$ diagram these objects lie close to each other, so it is possible that they are of the same nature.

\begin{figure}[t]
\centerline{\epsfysize=8.5cm
\epsfbox{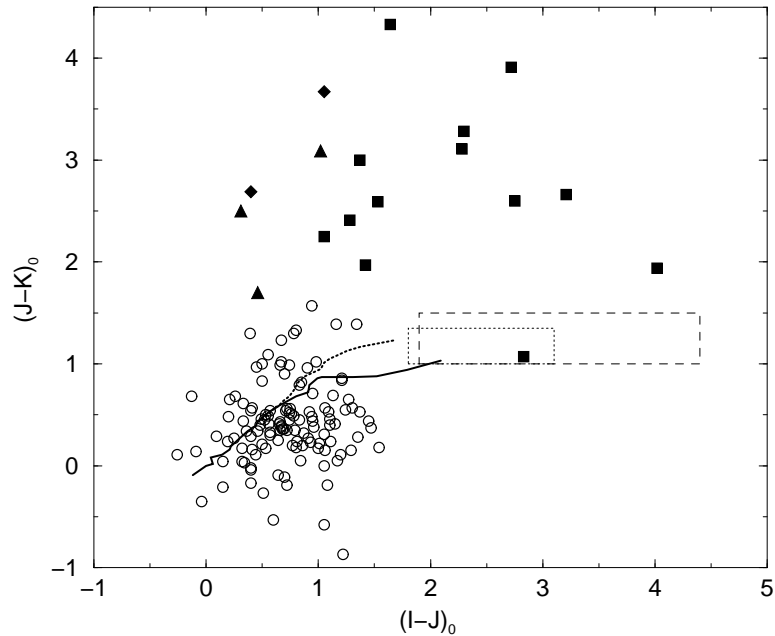}}
\caption{$IJK$ diagram of genuine PNe (circles), symbiotic Miras (squares), IR-[WC] PNe (triangles), and peculiar objects (diamonds).
Also shown are the positions of the stellar main sequence 
(solid line), the giants (dotted line), and the Miras and semiregular variables (dashed and dotted box, respectively; after Hron \& Kerschbaum 1994).}
\end{figure}

\end{document}